\documentclass[twoside]{article}
\usepackage{fleqn,espcrc2}
\usepackage{graphicx}

\newcommand{\AmS}{{\protect\the\textfont2
  A\kern-.1667em\lower.5ex\hbox{M}\kern-.125emS}}
\usepackage{graphicx}%style pour inserer figures en eps
\usepackage{here}%pour inserer figures et tables a emplacement 
\def\beq{\begin{equation}}
\def\eeq{\end{equation}}
\def\bea{\begin{eqnarray}}
\def\eea{\end{eqnarray}}
\def\bq{\begin{quote}}
\def\eq{\end{quote}}

\def\bear{\begin{array}}
\def\ear{\end{array}}
\def\nnb{\nonumber}
\def\ga{\left(}
\def\dr{\right)}

\def\rar{\rightarrow}
\def\Lrar{\Longrightarrow}

\def\nnb{\nonumber}

\def\nin{\noindent}
\def\ba{\begin{array}}
\def\ea{\end{array}}

\def\b{\bullet}

\def\gam5{\gamma_5}

\title{\bf{UV and IR divergences within Dimensional
Regularization in Non-Commutative theories and Phenomenological Implications}}
\author{ Stephan Narison\address{
Laboratoire de Physique Math\'ematique,
Universit\'e de Montpellier II
Place Eug\`ene Bataillon,
34095 - Montpellier Cedex 05, France. 
\\ E-mail:
qcd@lpm.univ-montp2.fr}}
\begin{document}
\pagestyle{empty}
\begin{abstract}
\noindent
Using Dimensional Regularization (DR) for some two-point functions of a prototype Non-Comutative (NC) $\phi^4$ scalar
theory in 4-dimensions, we explictly analyze, to one-loop, the IR and UV divergences of non-planar diagrams having
quadratic divergences and compare to the
case of the Pauli-Villars cut-off regularization (PVR). We also note that the IR structure
$1/p\circ p$ obtained from DR is reproduced by PVR in the limit where the UV cut-off
$\Lambda$ is set to
infinity. We study the phenomenological implications of this result by rederiving bounds from low-energy data on
the violation of Lorentz invariance based on the existence of the quadratic divergence. The most stringent (and
regularization independent) bound on Lorentz violation from low-energy data is $1/\sqrt{\theta}\approx \nu\geq
10^{15}$ GeV for NCQCD and $10^{10}$ GeV for NCQED, which comes from the absence of sidereal variations between the Cs and Hg
atomic clocks.
\vspace*{2mm}
\noindent
\end{abstract}
\maketitle
%%%%%%%%%%%%%%%%%%%%%%%
\section{Introduction}
\nin
At present, one expects that our current understanding of space-time may be modified at a very
short  distance scale.  One possible modification, inspired by quantum mechanics and 
also motivated by string theory arguments, is that the space-time coordinates
become noncommutative \cite{WITTEN}. The characteristic scale $\Lambda$ of the NC models, above which one may expect a
modification of the Standard Model, can
be parametrized by an ``angle" $\theta$ defined from the $x$-space commutation relation:
\begin{equation}
[\hat X_\mu, \hat X_\nu] = i \theta_{\mu\nu}~, ~~~{\rm where:}~~~\theta_{\mu\nu}\sim \frac{1}{\Lambda^2}~.
\end{equation}
Though it can be premature to perform an explicit calculation before a complete understanding of the
geometrical and mathematical foundations of the models and of its connection with the physical world, it
is interesting to check the self-consistency of the existing calculations for a given framework.
%\cite{loop}. 
In particular, one of the peculiar features of the NC approach is the correlation between the
UV and IR divergences of the theories, explicitly shown in some scalar models \cite{SEIBERG},
using the Pauli-Villars cut-off regularization (PVR). In this short note, we test, if the correlation of the UV and IR
divergences is, whether an articaft of PVR, or a more general phenomena independent of the regularization procedure. In so doing, we use
't Hooft-Veltman \cite{THOOFT} Dimensional Regularization (DR) \cite{SNB}, which is known to be a powerful method in gauge theories as it
implictly preserves gauge invariance and avoids quadratic divergences.
%%%%%%%%%%%%%%%%%%%%%%%%%%%%%%%%%%%%%%%%%%
\section{Scalar two-point functions in $\phi^4$ theory}
\nin
The lagrangian density of the NC $\phi^4$ theory is:
\beq
{\cal L}(x)={1\over 2}\ga\partial_\mu\phi\dr^2+{1\over 2}m^2\phi^2+{g^2\over 4!}\phi\star\phi\star\phi\star\phi~,
\eeq
where the $\star$ product is defined as:
\beq
\ga\phi_1\star\phi_2\dr(x)=e^{{i\over 2}\theta_{\mu\nu}\partial^y_\mu\partial^z_\nu}\phi_1(y)\phi_2(z)|_{y=z=x}~.
\eeq
Let's consider the 1PI two-point function, which, to lowest order, corresponds to the inverse propagator:
\beq
S^{(0)}(p^2)=p^2+m^2~.
\eeq
%%%%%%%%%%%%%%%%%%%%%%%%%%%%
\begin{figure}[hbt]
\begin{center}
\includegraphics[width=9cm]{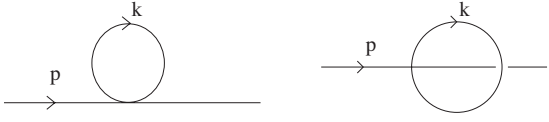}
\caption{\footnotesize{Planar and nonplanar one loop corrections to the inverse propagator in
$\phi^4$ theory.}}
\label{planar}
\end{center}
\end{figure}
\nin
%%%%%%%%%%%%%%%%%%%%%%%%%%%%
In NC theory, the one-loop corrections come from the planar and non-planar 
diagrams shown in (Fig. \ref{planar}), which lead to the corrections in $n$-space-time dimensions:
\bea\label{eq:one-loop}
&S^{(1)}_{planar}={g^2\over 3}\int{d^nk\over (2\pi)^{n}}{1\over k^2+m^2}~,\nnb\\
&S^{(1)}_{non-planar}={g^2\over 3!}\int{d^nk\over (2\pi)^{n}}{1\over k^2+m^2}e^{ik\times p}~,
\eea
where the NC-phase argument is:
$
k\times p\equiv k_\mu\theta^{\mu\nu}p_\nu~.$ It is usual in NC-calculations to parametrize the propagator \`a la
Schwinger:
\beq
{1\over p^2+m^2}=\int_0^\infty dz ~e^{-z(k^2+m^2)}~,
\eeq
while we use in $n$-dimensions the property:
\beq
\int{d^nk\over (2\pi)^{n}}e^{-z k^2}={1\over (4\pi z)^{n/2}}~.
\eeq
Using the previous inputs and the definition of the Gamma-function, the evaluation of the integrals in Eq.
(\ref{eq:one-loop}) is straightforward. 
$\b$ {\it The planar diagram}\\
The planar diagram gives for $n=4-\epsilon$:
\bea\label{eq:planar}
S^{(1)}_{planar}&=&\frac{g^2}{
48\pi^2}m^2\Gamma(\epsilon/2)\ga {m^2\over\nu^2}\dr^{\epsilon/2}\nnb\\
&\simeq&\frac{g^2}{
48\pi^2}m^2\Big{[}
{2\over\epsilon}-\log{m^2\over\nu^2}+\cdots\Big{]}~,
\eea
where $\nu$ is the typical UV scale of dimensional regularization, and $\cdots$ indicate regular terms. As expected,
the quadratic divergence obtained in the PV cut-off scheme is absent in DR. In effective theories, like e.g. the
well-known chiral perturbation theory of QCD, the $m^2\log(m^2/\nu^2)$ is a typical one-loop correction induced by a pion
loop with a mass $m$, while $\nu$ is fixed to be about the hadronic scale of about 1 GeV, where beyond this value the
effective approach is expected to be not valid.
\\ $\b$ {\it Non-planar diagram: $p\circ p\rar 0$ after integration}\\
The evaluation of the non-planar diagram is slightly more involved, but the
strategy is very similar. One obtains in
$n$-dimensions:
\beq\label{eq:nplanar}
S^{(1)}_{non-planar}={g^2\over 96\pi^2}\int_0^\infty {dz\over z^{n/2}}~e^{-z
m^2-{p{\small\circ} p\over 4z}},
\eeq
where:
$
p{\circ} q\equiv -p^\mu \theta^2_{\mu\nu}q^\nu=|p_\mu\theta^2_{\mu\nu}q_\nu|~.
$
We use the prescription where $n=4-\epsilon$ (4 is the space-time dimension and $\epsilon\rar 0$). The evaluation of this
integral in
$n\equiv 4-\epsilon$ dimension leads to:
\bea
S^{(1)}_{non-planar}&\simeq&{g^2\over 96\pi^2}2m^2\ga {4m^2\over p\circ p}\dr^{(1/2-\epsilon/4)}\nnb\\
&\times&K(1-\epsilon/2,\sqrt{m^2 p\circ p})~.
\eea
Expanding the Bessel function for $p\circ p\rar 0$:
\beq
K(1-\epsilon/2,~y\rar 0)\simeq {1\over y}+{y\over 2}\log y+{\cal O}\ga \epsilon,~y\dr~,
\eeq
one obtains for $\epsilon\rar 0$:
\beq
S^{(1)}_{non-planar}\simeq{g^2\over 24\pi^2}\Bigg{[}{1\over p\circ p}+{m^2\over 2}\log{m^2p\circ p}\Bigg{]}.
\eeq
This expression demonstrates that the non-planar diagram is UV finite in the $\epsilon\rar 0$ limit, i.e. in 4-dimensions,
and there is no need for introducing an UV cut-off. 
One may also deduce from this expression that the UV ($\epsilon\rar 0$) limit and the IR ($p\circ p\rar 0$) one  are 
independent to this order of perturbation theory. This result has to be contrasted with the one:
\beq\label{eq:pv}
S^{(1)}_{non-planar}\simeq{g^2\over 96\pi^2}\Bigg{[}\Lambda^2_{eff}+m^2\log\ga{m^2\over\Lambda^2_{eff}}\dr\Bigg{]},
\eeq
obtained using the PV cut-off scheme \cite{SEIBERG} in 4-dimensions, where
\beq\label{eq:pvb}
\Lambda^{-2}_{eff}\equiv\Lambda^{-2}+p\circ p~.
\eeq
Indeed, in Eqs. (\ref{eq:pv}) and (\ref{eq:pvb}), taking the UV ($\Lambda\rar\infty$) or/and IR ($p\circ p\rar
0$) limits
is ambiguous due to the mixing of the UV and IR divergences in the quadratic terms. The results of the DR and PV regularization
schemes for the IR divergence co\"\i ncide in the limit where the UV cut-off $\Lambda\rar\infty$. One should also notice that,
in this limit, both regularization schemes lead to a IR $1/p\circ p$ pole from the non-planar diagrams specific of the
NC-theories. This pole differs from the non-analytical
$\log (p\circ p)$ pole present in calculations having only logarithmic UV divergences using PV-cut'off regularizations, such as
the renormalization of the coupling constant $g^2$ in NC $\phi^4$ theory in four-dimensions, the vertex corrections in $\phi^3$
scalar theory in six-dimension...which have been  explicitly discussed in \cite{SEIBERG}. This feature presumably indicates
that the DR regularization though (apparently) decoupling the UV and IR divergences and transforming the UV quadratic divergence
into a logarithmic one, does not affect the structure of the IR $1/p\circ p$ pole from non-planar diagrams specific of the
NC-theories. 
\\ $\b$ {\it Non-planar diagram: $p\circ p$=0 before integration:}\\
However, one may also ask the question whether the operation taking the limit $\epsilon=0$ and $p\circ p=0$ is commutative
\footnote{Related questions have been addressed by the referee and by \cite{HUFFEL} in a slightly different form. However,
contrary to the claim of Ref. \cite{HUFFEL}, the result is not related to the fact of taking the intermediate step $n=2$
before analytically continuing to $n=4-\epsilon$ dimension, but only by taking first $p\circ p$=0 before integration.}.
Therefore, we put
$p\circ p=0$ in Eq. (\ref{eq:nplanar}) before integration. We obtain:
\beq\label{eq:nonplanar3}
S^{(1)}_{non-planar}\simeq{g^2\over 96\pi^2}(m^2)^{1-\epsilon/2}\Gamma(-1+\epsilon/2)~,
\eeq
which has now a $1/\epsilon$ pole \footnote{In the case of massless integral, we do not have such a pole if we use the standard
prescription that a tadpole-like integral $\int d^nk/k^2=0$ within dimensional regularization.}. This result indicates that the
$p\circ p$ term also acts (in an indirect way) as a regulator of the UV integral, which may be another aspect of the UV/IR
mixing within dimensional regularization. However, the question is to know if this second case is physically interesting~?
Indeed, taking
$p\circ p=0$ inside the integral is equivalent to reduce the non-planar into the planar one, and the result in Eq.
(\ref{eq:nonplanar3}) is nothing else (modulo normalization) than the planar one in Eq. (\ref{eq:planar}).
%%%%%%%%%%%%%%%%%%%%%%%%%%%%%%%%%%%%%
\section{Resummation of higher loops in $\phi^4$ scalar theory}
%%%%%%%%%%%%%%%%%%%%%%%%%%%%
One may generalize the previous one-loop result by resumming infinite series of divergent non-planar graphs including planar
one-loop mass corrections. This yields to \cite{SEIBERG}:
\beq
{\cal I}=\int{d^nk\over S^{(2)}(k)}~,
\eeq
with:
\beq
S^{(2)}(k)=m^2_R+k^2+{g^2\over 24\pi^2 k\circ k}+\cdots~,
\eeq
where $m_R$ is the renormalized scalar mass. The integral is IR finite and has the typical $1/\epsilon$ UV divergence for
$n=4-\epsilon$ space-time:
\beq
{\cal I}\sim m_R^2\ga{m_R^2\over 4\pi\nu^2}\dr^{-\epsilon/2}\Gamma(\epsilon/2)~.
\eeq
However, the extension of this resummed result to NC-Yang-Mills theory is not straightforward due to the presence of tachyonic
mass implying that the series does not have alternating signs \cite{RUIZ}. 
\begin{figure}[hbt]
\begin{center}
\includegraphics[width=7cm]{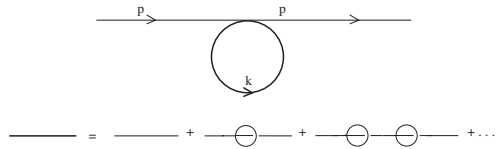}
\caption{\footnotesize{An infinite series of divergent graphs sums up to a single
graph with the dressed one loop propagator.}}
\label{planar}
\end{center}
\end{figure}
\nin
%%%%%%%%%%%%%%%%%%%%%%%%%%%%
%%%%%%%%%%%%%%%%%%%%%%%%%%%%%%%%%%%%%%%%%%%%
\section{Some phenomenological implications}
\nin
The absence of quadratic divergences within dimensional regularization can (a priori) affect different phenomenological
constraints on NC-models based on the existence of this term. In general, in a theory with a massive particle $m$, one should do
the replacement:
\beq\label{repl}
\Lambda^2\rar m^2\log {\nu^2 \over m^2}~,
\eeq
where $\nu$ is the UV scale of DR and corresponds to the scale until which the effective theory approach is expected to be valid.
This log-term is also present in PVR but non-leading compared with the $\Lambda^2$ one and then is regularization independent. 
Using this result, one expects that most of the stringent bounds on the scale of spacetime noncommutativity derived from the
low-energy tests of Lorentz violation, and which are based on the quadratic term are regularization dependent. 
Regularization independent constraints should then come from the $\log$ term. \\ $\b$ In the case of NCQCD \footnote{A
consistent construction of NCQCD has been proposed recently in \cite{WESS}.}, the magnetic operator
$\bar q\theta_{\mu\nu}\sigma^{\mu\nu}q$ acts like a $\vec\sigma\cdot \vec B$ interaction with a fixed $\vec B$ and leads to
sidereal variations in e.g. the hyperfine splittings \footnote{Some other low-energy constraints have been derived in
\cite{CHAICHIAN}, while constraints from high-energy accelerators have been e.g. derived in \cite{RIZZO}. We plan to re-examine
these existing constraints in a future publication
\cite{KING}.}. Such variations in the differences between Cs and Hg atomic clocks sensitive to external
$\vec\sigma\cdot
\vec B$ interactions are bounded at the
$10^{-31}$ GeV level. The authors in Ref.
\cite{CARLSON}, have estimated this operator in a nuclear environnment using a constituent quark model and \cite{POPSELOV} QCD
spectral sum rules (QSSR) approach
\cite{SNB}. Ref. \cite{CARLSON} obtained:
\beq\label{constr}
\ga{\alpha_s\over 12\pi} \dr p_0\Lambda^2\theta\leq \Delta E\Lrar \theta\Lambda^2\leq 10^{-29}~,
\eeq 
where $p_0\approx m_{con}\approx 300$ MeV is the off-shell light quark momentum; $\alpha_s\approx 1$ and $\Delta E$ is the bound
on the sidereal variation. 
%The numerical constraint looks a priori inconsistent as
%$\theta$ is expected to be of the order of
%$1/\Lambda^2$, such that the inequality would lead to $1\leq 10^{-29}$ ! 
Using our previous result in Eq. (\ref{repl}),
the constraint in Eq. (\ref{constr}) becomes:
\beq
\theta~ m^2_{con}\log{\nu^2 \over m^2_{con}}\leq 10^{-29}~,
\eeq
which leads to the tight lower bound on the scale of NCQCD:
\beq\label{bound}
1/\sqrt{\theta}\approx \nu\geq 10^{15}~{\rm GeV}~.
\eeq
The numerical value is apparently not affected by the choice of the regularization schemes. However, one should note that,
in the PVR, one has taken the cut-off $\Lambda$ to be about 1 TeV \cite{CARLSON} for deriving the constraint on $\theta$.
As this bound is stringent, it is also necessary to check the reliability of the QSSR result (effect of choice of the
nucleon operators, stability,...) and to use alternative tests such as lattice calculations.\\
$\b$ In the case of NCQED, the effect of a similar operator has been analyzed in \cite{BANKS}. In the case of the electron and
light quark constituents, the bound has typically the size:
\beq
\theta\Lambda^2\leq 10^{-19}~.
\eeq
The one from the electron is expected to be much weaker due to the electron mass suppression. Using again our result in Eq.
(\ref{repl}), one obtains for constituent light quarks:
\beq
\theta~ m^2_{con}\log{\nu^2 \over m^2_{con}}\leq 10^{-19}~,
\eeq
which leads to the lower bound on the scale of NCQED:
\beq\label{bound2}
1/\sqrt{\theta}\approx \nu\geq 10^{10}~{\rm GeV}~.
\eeq
This bound is again much stronger than the one from collider data. 
%%%%%%%%%%%%%%%%%%%%%%
\section{Conclusions}
\nin
We have explictly shown that, to one-loop, the type of correlation between the UV and IR divergences of the NC models
encountered using the Pauli-Villars cut-off regularization (PVR) procedure is not present or is different within dimensional
regularization (DR) due to the absence of quadratic divergence. We also note that the IR structure obtained from DR is
reproduced by PVR in the limit where the UV cut-off
$\Lambda
\rar \infty$. We plan to check if the previous conclusions continue to hold for some other processes
\cite{KING}, and to higher orders of PT series.
Finally, we have studied the phenomenological consequences of our result, by revising some bounds on the NC-theories based on the
existence of the quadratic divergences. The most stringent bound for the NCQCD scale derived from low-energy data is given in Eq.
(\ref{bound}), while the one for the NCQED is in Eq. (\ref{bound2}).
%%%%%%%%%%%%%%%%%%%%%%%%%%%
\section*{Acknowledgements} 
\nin
This work has been initiated when the author has visited the NCTS-Hsinchu and NTU-Taipei (Taiwan) in 2001.
It is a pleasure to thank
Luis Alvarez-Gaum\'e for various email communications, and
George Zoupanos for stimulating the publication of this note.  
%\vfill\eject
%%%%%%%%%%%%%%%%%%%%%%%%%%%%%%%%%%%


\begin{thebibliography}{99}
\bibitem{WITTEN}N. Seiberg and E. Witten, hep-th/9908142.
\bibitem{SEIBERG}S. Minwalla, M. Van Raamsdonk and N. Seiberg, hep-th/9912072.
\bibitem{THOOFT}G. 't Hooft and M. Veltman, Diagrammar, CERN yellow report (1973); G. Leibbrandt, {\it Rev. Mod. Phys.} {\bf 47}
(1975) 849.
\bibitem{SNB}For reviews, see e.g.: S. Narison, {\it Phys. Rept.} {\bf 84} (1982) 263; QCD as a theory of hadrons, 
{\it Cambridge Monogr. Part. Phys. Nucl. Phys. Cosmol.} {\bf 17} (2002) 1 [hep-ph/0205006];
{\it World Sci. Lect. Notes Phys.}  {\bf 26}, 1 (1989).
\bibitem{HUFFEL}H. H\"uffel, hep-th/0210028.
\bibitem{RUIZ}F. Ruiz Ruiz, hep-th/0012171. 
\bibitem{WESS} B. Jur$\tilde c$o et al, hep-ph/0104153. 
\bibitem{CHAICHIAN}M. Caichian, M.M. Sheikh-Jabbari and A. Tureau, hep-th/0010175; S.M. Caroll et al. hep-ph/0106356; I.
Hinchliffe and N. Kersting, hep-ph/0104137.
\bibitem{RIZZO}J. Hewett, F. Petriello and T. Rizzo, hep-ph/001035.
\bibitem{KING}A. Arhib, K. Cheung and S. Narison (in preparation).
\bibitem{CARLSON}C.E. Carlson, C.D. Carone and R.F. Lebed, hep-ph/0107291.
\bibitem{POPSELOV} I. Moscioiu, M. Pospelov and R. Roiban,
hep-ph/0005191.
\bibitem{BANKS} A. Anisimov, T. Banks, M. Dine and M. Grasser, hep-ph/0106356.
\end{thebibliography}
\end{document}